# The FLUKA code: present applications and future developments

A. Fassò, A. Ferrari, S. Roesler
*CERN*

J. Ranft
*Siegen*

P.R. Sala
*ETHZ*

G. Battistoni, M. Campanella, F. Cerutti, L.De Biaggi, E. Gadioli, M.V. Garzelli
*Milano*

F. Ballarini, A. Ottolenghi, D. Scannicchio
*Pavia*

M. Carboni, M. Pelliccioni, R. Villari
*LNF*

V. Andersen, A. Empl, K. Lee, L. Pinsky
*Houston University*

T.N. Wilson, N. Zapp
*NASA/JSC*

The main features of the FLUKA Monte Carlo code, which can deal with transport and interaction of electromagnetic and hadronic particles, are summarised. The physical models embedded in FLUKA are mentioned, as well as examples of benchmarking against experimental data. A short history of the code is provided and the following examples of applications are discussed in detail: prediction of calorimetric performances, atmospheric neutrino flux calculations, dosimetry in atmosphere and radiobiology applications, including hadrontherapy and space radiation protection. Finally a few lines are dedicated to the FLUKA server, from which the code can be downloaded.

## 1. INTRODUCTION

FLUKA is a Monte Carlo code able to simulate transport and interaction of electromagnetic and hadronic particles in any target material over a wide energy range [1-4]. It is a multi-purpose, multi-particle code that can be applied in many different fields. Particular attention has been devoted to following the various components of the hadronic and electromagnetic cascades. Thus the code characteristics at intermediate energies make it particularly reliable in treating problems in the fields of radiotherapy and radiation protection.

As in most simulation codes that adopt a "condensed-history" approach, in FLUKA "continuous" processes such as energy loss and angular deflections due to Coulomb interactions and "discrete" (or "explicit") processes (delta-ray production, nuclear interactions, decays, bremsstrahlung and photon interaction) are treated separately. A detailed description of the models implemented in FLUKA can be found elsewhere [1-3], and only the main features will be summarized herein.

Energy loss by charged particles is described on the basis of the Bethe theory, supplemented with average ionisation potentials, density and shell corrections according to ICRU publications no. 37 and 49 [5]. The description of multiple Coulomb scattering relies on Moliere's theory. A detailed algorithm has been developed to simulate the path curvature effects. The production of delta-rays is described explicitly above a user-defined threshold (e.g. 10 keV for radiobiology applications), below which a continuous energy loss with statistical fluctuations is assumed. Concerning anelastic nuclear interactions at intermediate energies, FLUKA uses the PEANUT (Pre-Equilibrium Approach to NUclear Thermalisation) model, which consists of intranuclear cascades (INC), pre-equilibrium, evaporation and de-excitation. In case of light nuclei, for which the evaporation model is not very significant, the Fermi break-up model has been implemented. At energies above 5 GeV/n, a model based on the Dual Parton Model (DPM) is used.

The current version of the code can simulate electromagnetic and μ interactions up to 100 TeV, neutron interaction and transport down to thermal energies (multigroup below 20 MeV) and hadron-hadron and hadron-nucleus interactions up to 100 TeV. Nucleus-nucleus interactions, previously available above 5 GeV/n only (up to 10,000 TeV/n), have been recently implemented also down to ≈100 MeV/n by coupling FLUKA to the Relativistic Quantum Molecular Dynamics code RQMD-2.4 [6].

FLUKA can treat both combinatorial geometry and voxel geometry, and an interface to GEANT4 geometry is also available. Presently the code is maintained for various platforms with Unix Interface: Linux, Compaq-Unix, HP-Ux and Sun-Solaris.

The validity of the physical models implemented in FLUKA has been benchmarked against a variety of experimental data over a wide energy range, from accelerator data to cosmic ray showers in the Earth atmosphere. In particular, a vast amount of benchmarking against muon, hadron and electron data in atmosphere [7-9] has confirmed the accuracy of FLUKA hadronic interaction models.

FLUKA is widely used for studies related both to basic research and to applications in radiation protection and dosimetry, radiobiology (including radiotherapy) and cosmic ray calculations. Two recent, important examples of calculation related to cosmic rays are the calculation of atmospheric neutrino fluxes [10] and the evaluation of aircraft exposure [9,11]. A modified version of the code has been purposely developed for radiobiological studies, mainly for applications in hadrontherapy and space radiation protection. More specifically, results of track structure simulations at the nanometre level have been integrated on-line in FLUKA, thus allowing calculation of "biological" doses (see below). The integration method has been applied to the characterisation of therapeutic proton beams [12,13] and to the calculation of organ doses in case of exposure to the Solar Particle Event component of space radiation [14,15]. The latter has been performed by coupling the code with two anthropomorphic phantoms, a mathematical model and a voxel model.

## 2. A SHORT HISTORY OF FLUKA

The FLUKA code started being developed in 1962 by J. Ranft and H. Geibel, who initiated the code for hadron beams. The name FLUKA (from FLUktuierende KAscade) came eight years later, since at that time the code was mainly used for applications concerning event-to-event fluctuations in calorimetry. Between 1970 and 1987 the development of the code was carried out in the framework of a collaboration between CERN and the groups of Leipzig and Helsinki. That version was essentially for shielding calculations.

Since 1989 FLUKA is being developed within INFN (National Institute of Nuclear Physics) with the personal collaboration of A. Fassò (CERN) and J. Ranft (Leipzig). One of the main aims is developing an all-purpose, general code with new physics models. Presently very little was left of the 1987 version.

In 1990 MCNPX officially started using FLUKA for its high energy part. In 1993 FLUKA was interfaced to GEANT3 (for the hadronic part only). This interface did not follow the subsequent FLUKA developments and it is therefore now obsolete.

Since 2002 FLUKA is an INFN project, with the main aim of providing a better diffusion of the code and stimulating all those studies that can be of interest for applied research, with focus on dosimetry, medical physics and, more generally, radiobiology. The INFN project is carried on in strict collaboration with CERN and the University of Houston. Further information can be found in the FLUKA web site (http://www.fluka.org).

Finally it is worth mentioning that in 2003 a joint INFN-CERN project was initiated, with the aim of developing, maintaining and distributing the FLUKA code.

## 3. APPLICATIONS OF FLUKA

### 3.1. High Energy Applications

Within the framework of high-energy physics studies, FLUKA was applied to the prediction of calorimetric performances [16]. More specifically, the pion resolution expected for the ICARUS detector was calculated, and predictions relative to the ATLAS calorimeter set-up were compared to experimental data.

The ICARUS collaboration has developed a technique based on the fact that ionisation electrons can drift over long distances within a volume of purified liquid Argon under a strong electric field, that allows obtaining high quality 3-D imaging. Since the medium appears as a completely homogeneous volume with very high readout granularity, the event visualisation and the local charge deposition density allow one to distinguish between the electromagnetic and the hadronic component of a shower, as well as to approximately correct for recombination effects. By assuming that each elementary cell contains only one crossing track, the recombination effect can be unfolded using the collected charge and the cell width to construct the observed dQ/dx, and solving the recombination expression for the "actual" dE/dx. In order to reconstruct the energy, assuming that the electromagnetic contribution $Q_{em}$ can be distinguished from the hadronic contribution $Q_{had}$, the total energy E of a shower can be calculated as $E=w(Q_{em} + \alpha Q_{had})$, where $\alpha$ is the compensation factor. On these bases, the expected hadronic resolution for pions in the ICANOE detector was calculated under different conditions (with and without TMG doping, with and without compensation, with and without quench corrections) and compared with the one that would be obtained with no recombination effect.

The ATLAS electromagnetic and hadronic calorimeter have been tested with $\pi$, $\mu$ and electron beams from 10 to 300 GeV/c. The experimental set up was reproduced with FLUKA and the simulation results were compared with $\pi$ and $\mu$ data, by applying the same cuts as in the experiments. Charge collection and signal quenching were simulated, whereas electronic noise and photo-statistics were included *a posteriori*. Proton contamination in the pion beam was also taken into account. The results obtained with FLUKA were calibrated in electron scale without any further normalisation. The visible energy $E_0$ was reconstructed with the "benchmark" technique:

$E_0 = E_{em} + aQ_{had} + b|E_{em3}\, a\, Q_{had1}|^{1/2} + cE_{em}^2$

All parameters were fixed to minimise $\sigma/E_0$ at 300 GeV. The parameter values predicted with FLUKA were in very good agreement with the measured ones, and the non-compensation as a function of energy was well reproduced. The fractional energy resolution was also calculated. Although the constant and noise terms were slightly underestimated, the sampling term was well reproduced.

### 3.2. Cosmic Ray Applications

FLUKA has been successfully applied in cosmic ray physics for the analysis of experimental data and the calculation of secondary particle fluxes in atmosphere. Among the various pieces of work it is worth while mentioning the first successful 3-D calculation of atmospheric neutrino fluxes [17]. This task was initiated in the framework of the ICARUS [18] and MACRO [19] experiments at the INFN Gran Sasso laboratories.

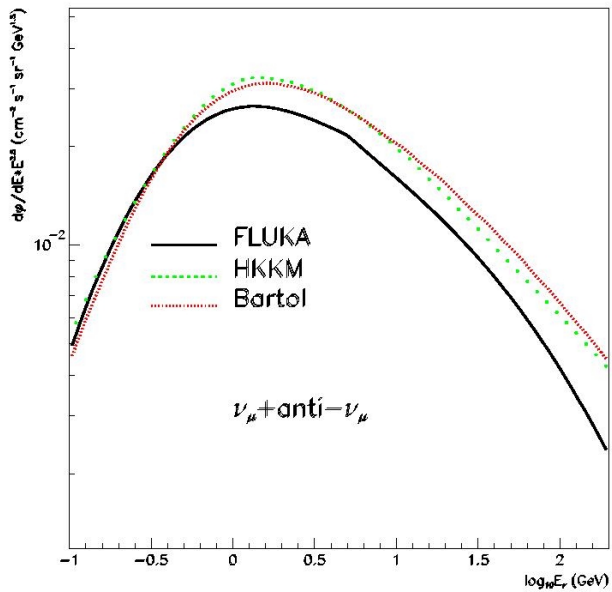

Figure 1: Calculated atmospheric muon neutrino plus antineutrino flux, angle averaged over the lower Earth hemisphere as a function of neutrino energy. The FLUKA calculation is compared to other models.

One of the main merits of this work was the stimulation of a deep check of the quality of hadronic physics modeling and of its impact in this line of phenomenological research. In connection to the topic of atmospheric neutrino fluxes, the FLUKA interaction model has been used also for other secondary particles produced in atmosphere by cosmic rays, which can be used as a further cross check of the validity of the model. At least two remarkable results can be quoted. The first is the reproduction of the primary proton flux features as a function of geomagnetic latitude as measured by AMS [20], which showed that the geomagnetic effects and the overall geometrical description of the 3-D setup are well under control [8]. The same work showed that also the fluxes of secondary $e^+e^-$ measured at high altitude are well reproduced. The second is the good reproduction of muon data in atmosphere as measured by the CAPRICE experiment [21], both at ground level and at different floating altitudes: see ref. [7] for the relevant plots and numbers. The agreement shown by the FLUKA simulation for muons of both charges gives confidence on the predictions of FLUKA for the parent mesons of muons (mostly pions). This work complements the previously mentioned studies, oriented to the validation of the model in terms of particle yields.

The shower simulations in atmosphere with FLUKA were compared also to the most recent hadron spectra at different latitudes and altitudes, obtaining remarkable agreement, as for example in the case of the hadron flux measured with the calorimeter of the KASCADE experiment [22]. Thanks to these results, the hadronic part of FLUKA will be interfaced to the CORSIKA code [23] (widely used in high energy cosmic ray physics) as an alternative to GHEISHA for the interactions below 100 GeV. Concerning future developments, the inclusion of the DPMJET model for ion interaction at high energies will also allow to fully simulate cosmic ray interactions up to $10^{20} - 10^{21}$ eV. As a complement to this topic, we also can quote valuable simulations of the features of extensive air showers [24] and also of the muon transport in the rock [25], which was an important step for the data analysis of underground experiments like MACRO, also to understand the local hadron production by muon in the underground environment [26], thus allowing a reliable study of neutron background for experiments searching for rare events and dark matter.

The results described above provided a starting basis for applications related to dosimetry in atmosphere, which will be described in the next section.

### 3.3. Dosimetry in atmosphere

Cosmic radiation penetrating into the Earth's atmosphere produces a very complex environment. The largest fraction of radiation encountered by commercial jet aircrafts is the secondary radiation produced when galactic cosmic rays interact with the nuclei of the air constituents. Events caused by sporadic eruptions of the Sun chromosphere will occasionally result in unforeseen additional contribution to the radiation field (solar flares). According to the ICRP Publication 60 [27] and to the national laws incorporating the new European Directive (96/29/Euratom), the exposure of air crews to cosmic radiation in jet aircrafts should be considered as an occupational exposure. For air crew whose annual dose falls in the range 1-6 mSv individual dose estimates should be required. For flights below 15 km such estimates may be carried out using appropriate computer programs. Particles penetrating the atmosphere generate a cascade consisting of the hadronic, the electron-photon and the muonic component. The intensities of the various components are predictable with the exceptions of rare solar events. Dose rates vary with the depth in the atmosphere, the geomagnetic latitude and, less markedly, with the phase of the solar cycle.

To determine a simple procedure for a realistic dose assessment at aviation altitudes, calculations of atmospheric showers initiated by galactic cosmic rays were carried out using FLUKA. Primary spectra used as inputs in the calculations corresponded to values of the deceleration potential $\Phi$ of 465 and 1440 MV. Simulations were carried out for several values of vertical cut-off, $R_V$, from 0.4 GV to 17.6 GV. Further calculations were performed for $\Phi$ values of 600, 800, 1000 and 1200 MV, but only for vertical cut-off of 0.4 and 3.0 GV. Primary particles were isotropically generated at the top of the atmosphere with geomagnetic modulated spectra. The fluences of the various components produced in air were determined at different depths in atmosphere and at different geomagnetic latitudes. The details can be found in [28,29].

The quantities of interest in radiological protection, that is effective dose and ambient dose equivalent, were obtained by folding the calculated particle spectra with appropriate sets of conversion coefficients [30]. The geometrical conditions of irradiation were assumed to be isotropic. As an example, figure 2 shows the ambient dose equivalent rate at various altitudes, together with the contributions of the individual secondary particles. Although calculations are associated with large uncertainties, which comes from different sources (primary spectra, application of the superposition model etc.), a generally satisfactory agreement was found between the present calculation results and experimental data or results of other calculations.

Simple equations, relating the effective dose to the various parameters involved (latitude, altitude and solar cycle), have been proposed at civil aviation altitudes [29,31]. The use of the proposed equations for air crew dose assessment requires information about flight routes and altitude profiles. The key point of such a practical evaluation method is the fact that even those who have not access to sophisticated computer codes should be able to work out individual dose assessments.

It is worth being noted that the proposed equations, as well as the results of other computer codes, do not include the unpredictable contributions of solar particle events, as well as the effects due to the additional layers of the aircraft structures. In order to evaluate the influence of the aircraft structures on the dose values, a simplified geometry of an Airbus-340 was arranged. The model is constituted by 144 FLUKA regions and it includes 11 different materials to describe fuselage structure, fuel, passengers and cargo. The geometry was adjusted to allow for the modification of some parameters (i.e. number of passengers, load etc.). Figure 3 shows a view of the transversal and longitudinal sections of the aircraft.

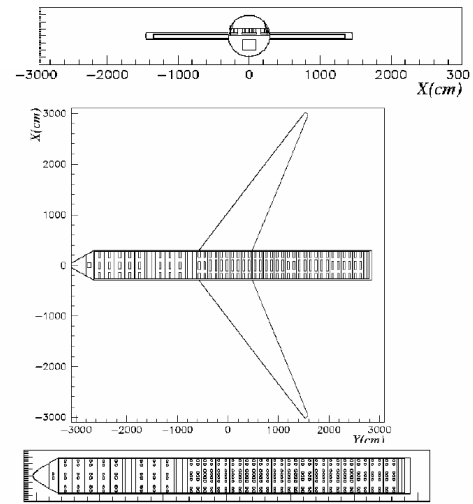

Figure 3: Sketch of the transversal and longitudinal sections of an Airbus-340 simulated with FLUKA.

### 3.4. Radiobiology Applications

#### 3.4.1. Biophysical characterisation of therapeutic hadron beams

Tumor treatment with hadrons (mainly protons, but also Carbon ions) is becoming more and more widespread, and more than 30,000 patients have been treated up to now. This is mainly due to the hadron beam capability of delivering most of the dose within a localized region called "Bragg peak", which allows for a better conformation of the dose distribution to the target volume. However, there is still a strong need for an improved characterisation of therapeutic hadron beams, both from a physical point of view and from a biological one. In particular the role of nuclear reaction products, which can have higher biological effectiveness with respect to the primary particles, needs to be evaluated at a quantitative level. Furthermore, average quantities such as absorbed dose, LET (Linear Energy Transfer) and RBE (Relative Biological Effectiveness) do not take into account the stochastic aspects of energy deposition in matter. There is therefore a strong need for the identification of new parameters able to provide a direct correlation between radiation track structure at the nanometre level and radiobiological effects. Clustered – and thus severe – DNA damage has been suggested to play a fundamental role in the induction of cell death [32], as well as other endpoints such as chromosome aberrations [33], which are correlated both with cell death (important in radiotharapy) and with cell conversion to malignancy (important in radiation protection). In a previous work [34], "Complex Lesions" (CL) have been defined as "at least 2 breaks in each of the two DNA strands within 30 base-pairs". CL yields per unit dose and DNA mass have been calculated for different radiation types and energies by coupling a DNA model to an "event-by-event" code describing energy deposition at the nanometre level.

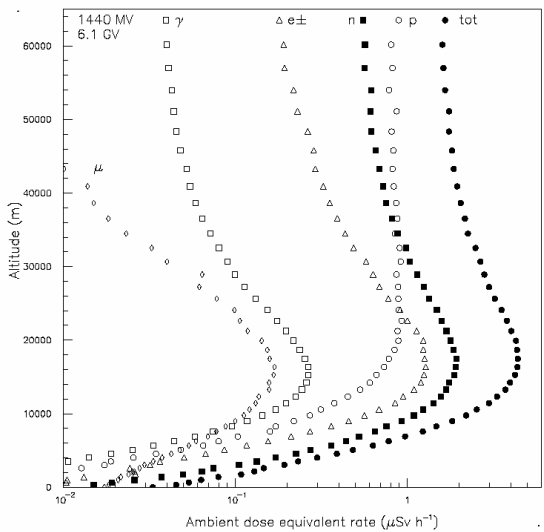

Figure 2: Ambient dose equivalent rate at various altitudes. The contributions of various individual secondary particles are also shown.

Since event-by-event simulations at a tissue/organ level would require unreasonable CPU times, an alternative approach was adopted, consisting of integrating CL yields in the condensed-history FLUKA code. This allowed us to calculate not only depth-dose profiles, but also distributions of the average number of CL/cell, a quantity that we called "biological" dose. In particular the integration method was applied to characterize the 72 MeV proton beam used for the treatment of ocular tumors at the Paul Sherrer Institute, Switzerland [12]. *Ad hoc* experiments for comparison were carried out at PSI: the therapeutic apparatus was set up in order to obtain a fully Spread-Out Bragg Peak (SOBP) in a perspex phantom and the depth-dose profile was measured, as well as survival of cells irradiated *in vitro* at different depths. In the simulations, the geometry of the apparatus was faithfully reproduced. The modulation of the beam, obtained at PSI with Aluminum profiles mounted on a rotating wheel, was simulated with a dynamical approach consisting of random mapping of a single profile along the beam axis. This allowed us to obtain a simulated Spread-Out Bragg Peak with a single run of the program. The depth-dose distribution calculated with FLUKA was found to be in excellent agreement with that measured at PSI. The contributions of the various beam components were calculated separately. The secondary hadron component accounted for less than 4% in most of the SOBP, and disappeared in its distal part. In contrast with the physical dose, which was roughly constant with depth throughout the SOBP, the calculated profile of the "biological" dose (i.e. average number of CL/cell) showed a sharp increase at the distal part of the SOBP, due to the presence of protons with low energy and thus high biological effectiveness. Furthermore, nuclear reaction products were found to play a more relevant role, reaching values around 12% in the proximal part of the SOBP. Excellent agreement was also found by comparing the (calculated) ratio between proton-induced CL and X-ray-induced CL to the (measured) ratio between proton-induced lethal lesions and X-ray-induced lethal lesions. The RBE was found to be 1.2 along most of the SOBP (except for the increase in the the distal part), consistent with the constant value of 1.1 used in therapeutic practice with proton beams.

Similar results were found by generalising the method described above to the characterisation of a 160 MeV proton beam, modulated in order to obtain a typical SOBP suitable for the treatment of deep solid tumours [13]. The calculated depth-dose profile in water was found to be constant along the SOBP. Secondary hadrons accounted for $\approx$6% of the total dose at the beginning of the plateau, $\approx$3% in most of the SOBP and disappeared at the distal part of the SOBP. By contrast, the distribution of "biological" dose showed a sharp increase at the distal part of the peak and a higher role of secondary hadrons ($\approx$20% at the beginning of the plateau and $\approx$15% in the SOBP). A RBE of $\approx$1.2 was found both along the plateau and in most of the SOBP, with a sharp increase in the distal part of the peak due to low energy protons. That the RBE was constant can be explained by taking into account that the slowing down of protons, which gives rise to high-LET particles, is balanced by the progressive decrease in the imporance of secondary hadrons along the plateau and the SOBP.

### 3.4.2 Space Radiation Protection

Missions onboard the International Space Station and a possible mission to Mars, which is in NASA's plans for the first half of our century, would imply exposure to high-energy protons and heavier ions constituting the spectra of Galactic Cosmic Rays and Solar Particle Events. While GCR are always present and are characterised by a maximum fluence rate of 4 particles cm$^{-2}$ s$^{-1}$, SPE are sporadic and unpredictable events with fluences up $10^{10}$ particles/cm$^2$ within a few hours or days. In both cases there is a strong need for shielding optimisation, a problem that can be approached with transport codes possibly coupled with human body models. In contrast with GCR, for which ions heavier than protons provide a fundamental contribution to the equivalent dose, the proton component of SPE accounts for $\approx$90% not only of the physical dose but also of the equivalent dose, and particles other than protons can be neglected in most cases. In this context, distributions of physical, equivalent and biological dose (in Gy, Sv and CL/cell, respectively) in the various organs of the human body were calculated by coupling the FLUKA code with two anthropomorphic phantoms [15]. The first one is a mathematical model in which the various organs are described with combinatorial geometry [35], whereas the second one is a "voxel" model described by more than 2 million voxels derived from total body CT data of a leukaemia patient [36]. No significant difference was found between the dose values calculated with the two phantoms. The phantoms were inscribed into an Aluminium box representing the shielding structure and were irradiated from an isotropically-emitting sphere of 2 m radius, external with respect to the phantom and the box. The thickness of the box was fixed at 1 and 2 $g/cm^2$ (nominal spacesuit), 5 $g/cm^2$ (nominal spacecraft) and 10 and 20 $g/cm^2$ (storm shelter to be used in case of SPE).

Irradiation was performed with the proton component of the two most intense SPE of the last 50 years, that is the October 1989 event and the August 1972 event. The latter has been estimated to be lethal for an unprotected crew on the Moon surface. The contributions of primary protons and secondary hadrons (including ions) produced in nuclear interactions with the shield and the human body were calculated separately. All types of dose (physical, equivalent and biological) were found to decrease with increasing shielding and the highest dose values were found for the skin. With the lowest shielding, the skin was found to receive 2.4 Gy, 3.5 Sv and 1.5 CL/cell for the 1989 event, and 7 Gy, 11.5 Sv and 4.4 CL/cell for the 1972 event, which was more effective due to the higher fluence in the energy range 20-200 MeV. While for the 1989 event 5 $g/cm^2$ Al were sufficient to respect the NCRP limits for 30 days missions in Low Earth Orbit (1.5 Gy-Eq for skin and 1 Gy-Eq for eye lenses, [37]), for the 1972 event 10 $g/cm^2$ were needed. The contribution of secondary hadrons to the physical dose was found to be smaller than 10% for all organs/tissues and for all the considered values of shielding thickness, except for some internal organs such as liver. As a general trend, the relative contribution of secondary particles was higher for internal organs with respect to skin and lenses, and increased with shielding thickness for a given organ or tissue. Similar trends (dose decrease with increasing shielding and higher contribution of secondaries for the internal organs) were found for the biological dose. However, from a quantitative point of view the role of nuclear interaction products was found to be higher for the biological dose with respect to the physical dose. The highest contributions were found for internal organs behind large shielding (e.g. 30% contribution for liver behind 10 $g/cm^2$ Al shielding in case of the 1989 event).

The role of nuclear reaction products is expected to become more important at higher energies and/or higher Z values. A preliminary test performed with a monochromatic high-energy proton beam (500 MeV) confirmed this hypothesis, showing that the contributions of secondary hadrons to the biological dose was even larger than that of primary protons [14]. The scenario will become even more complex when we will simulate the exposure to GCR spectra, for which heavier ions up to Fe play a relevant role. The simulation of GCR exposure, as well as the characterization of therapeutic ion beams (tipically Carbon), is now possible due to the recent implementation in FLUKA of nucleus-nucleus interactions below 5 GeV/n [4].

## 4. THE FLUKA SERVER

A major goal of the INFN FLUKA project is the dissemination of FLUKA and its evolution towards a better documented and relatively user-friendly code. For this purpose a fluka.org dominion was created, installing a server that is supported by the Italian scientific network (GARR) using the INFN network structure. By accessing the web server (http://www.fluka.org) it is possible to download the FLUKA code upon registration. At present, the users can access the FLUKA library, the user routine sources, the cross section data files (in binary form) and the essential scripts and tools for compilation, run and analysis, for RedHat Linux and different Unix platforms. In the future, according to the milestones of the INFN project and also to a recent CERN-INFN agreement for the development of FLUKA, the full source code will be accessible, under a suitable licensing scheme that is under study. The FLUKA web page is also providing access to the present version of the manual, which is still in evolution towards an html format, together with the possibility of having it as a pdf file. Furthermore, the web page is now used to provide a number of documented examples (evolving in time) which help the users to understand the practical utilisation of FLUKA, paying attention to the user requests received so far.

A FLUKA discussion list is also served and its archive is accessible from the web page, in order to provide a FAQ service. FLUKA history and references are also available. A series of FLUKA instruction courses is under study, with the possibility of providing, again through the web server, video recordings of the main lectures.

## Acknowledgments

This work was partially supported by the EC (contract no. FIGH-CT1999-00005, 'Low Dose Risk Models') and by the Italian Space Agency (contract no. I/R/320/02, 'Influence of the shielding on the space radiation biological effectiveness').